\documentclass[structabstract]{aa}
\usepackage{txfonts}
\usepackage{graphicx}
\usepackage{natbib}
\bibpunct{(}{)}{;}{a}{}{,}

\usepackage{multirow}

\def\kHz{ \textrm{kHz} }
\def\MHz{ \textrm{MHz} }
\def\GHz{ \textrm{GHz} }

\def\Hz{ \textrm{Hz} }
\def\K{ \textrm{K} }
\newcommand{\+}{$^+$}

\begin{document}

\title{HSCO\+ and DSCO\+: a multi-technique approach\\  in the laboratory for the spectroscopy of interstellar ions}

\author{Valerio Lattanzi\inst{1} \and Silvia Spezzano\inst{1} \and Jacob C. Laas\inst{1} \and Johanna Chantzos\inst{1} \and Luca Bizzocchi\inst{1} \and Kin Long Kelvin Lee\inst{2,3} \and Michael C. McCarthy\inst{2,3} \and Paola Caselli\inst{1}}

\institute{Max-Planck-Institut f\"ur extraterrestrische Physik, Giessenbachstra{\ss}e 1, D-85748 Garching, Germany  \and Harvard-Smithsonian Center for Astrophysics, Cambridge, Massachusetts 02138, USA \and School of Engineering and Applied Sciences, Harvard University, 29 Oxford St., Cambridge, Massachusetts 02138, USA\\
\email{lattanzi@mpe.mpg.de}}

\abstract{
Protonated molecular species have been proven to be abundant in the interstellar gas. This class of molecules is also pivotal for the determination of important physical parameters for the ISM evolution (e.g. gas ionisation fraction) or as tracers of non-polar, hence not directly observable, species. The identification of these molecular species through radioastronomical observations is directly linked to a precise laboratory spectral characterisation.
}{
The goal of the present work is to extend the laboratory measurements of the pure rotational spectrum of the ground electronic state of protonated carbonyl sulfide (HSCO\+) and its deuterium substituted isotopomer (DSCO\+). At the same time, we show how implementing different laboratory techniques allows the determination of different spectroscopical properties of asymmetric-top protonated species.}
{
Three different high-resolution experiments were involved to detected for the first time the $b-$type rotational spectrum of HSCO\+, and to extend, well into the sub-millimeter region, the $a-$type spectrum of the same molecular species and DSCO\+. The electronic ground-state of both ions have been investigated in the 273--405\GHz frequency range, allowing the detection of 60 and 50 new rotational transitions for HSCO\+ and DSCO\+, respectively. 
}{
The combination of our new measurements with the three rotational transitions previously observed in the microwave region permits the rest frequencies of the astronomically most relevant transitions to be predicted to better than 100 kHz for both HSCO\+ and DSCO\+ up to 500 GHz, equivalent to better than 60\,m/s in terms of equivalent radial velocity.
}{
The present work illustrates the importance of using different laboratory techniques to spectroscopically characterise a protonated species at high frequency. Each instruments addressed complementary part of the same spectroscopic challenge, showing that a similar approach can be adopted in the future when dealing with similar reactive species. 
}


\titlerunning{Protonated OCS}
\authorrunning{Lattanzi et al.}

\maketitle

\section{Introduction}

Carbonyl sulfide (OCS) has been observed in the ISM in several types of objects \citep{gol81, li15}, and it is the only S-bearing molecule so far unambiguously detected in interstellar ices \citep{2015ARA&A..53..541B}. Furthermore, it has been observed in the coma of the comet 67P \citep{2016MNRAS.462S.170B} and, recently, OCS has been mapped towards the pre-stellar core L1544 showing a spatial distribution comparable to methanol \citep{2017A&A...606A..82S}, possibly hinting at a common formation path onto the icy mantels of dust grains, as suggested by \citet{2012MNRAS.421.1476L}. In molecular clouds, OCS molecules could undergo protonation or deuteronation via reactions with H$_3^+$ and its deuterated isotopomers.

The presence of protonated OCS in the ISM has been suggested by several groups \citep{1982ApJ...257L..99F, tur90}. OCS possesses a large proton affinity (632\,kJ/mol), higher than either CO or CO$_2$, which are 576\,kJ/mol and 551\,kJ/mol, respectively \citep{2001PCCP....3.4359H}, whose protonated variants are known to exist in space \citep{2017A&A...602A..34B}. Radioastronomical observations of protonated carbon dioxide have been used in the past also to constrain the abundance of CO$_2$ in different interstellar environments. \citet{2016A&A...591L...2V} derived an indirect estimate of the [CO$_2$]/[CO] from [HOCO\+]/{HCO\+} in the L1544 pre-stellar core; similar arguments were used by \citet{2008ApJ...675L..89S} to study the Class 0 protostar IRAS 04368+2557 embedded in L1527. The Heterodyne Instrument for the Far Infrared (HIFI) mounted on the Herschel satellite was used by \citet{2014ApJ...789....8N} to observe submm/THz emission of HOCO\+ towards SgrB2(N), which, in conjunction with previously published observations of HCO+ isotopologues, allowed the authors to infer the gas-phase CO$_2$ abundance in the region. Protonated species and their deuterated variants can also give important information about the deuteration in interstellar clouds. DOCO\+ was searched in many environments where its parent species was also detected; new accurate rest frequencies from laboratory spectroscopic studies on DOCO\+ \citep{2017A&A...602A..34B} ruled out a tentative detection reported by \citet{2016A&A...591L...2V} towards the L1544 pre-stellar core.

According to the estimates by \citet{1982ApJ...257L..99F}, the molecular abundance in dense clouds of protonated carbonyl sulfide may be comparable to that of HOCO\+, which is also isovalent with HSCO\+. Carbonyl sulfide can be protonated both on the sulfur (HSCO\+) and the oxygen side (HOCS\+), with the former being more stable by $\sim$ 20 kJ/mol \citep{whe06}. 

Highly reactive molecules, such as ions and radicals, are difficult to produce and detect at high spectral resolution in the laboratory. Surprisingly, the higher--energy isomer, HOCS\+, was detected in the laboratory first: \citet{nak87} observed this cation with infrared spectroscopy with a hollow cathode discharge, followed by \citet{ohs96}, who further characterised its spectrum in the cm-wave range by means of a Fourier-transform microwave spectrometer coupled with a pulsed-discharge nozzle. Twenty years later, the rotational spectrum of the lower energy isomer, HSCO\+, along with its deuterium and $^{34}$S isotopic substituted species, was finally measured by \citet{mcc07}, using a similar experiment to that used by \citet{ohs96}. 

Rest frequencies derived from the current laboratory datasets are not accurate enough to confidently search for either isomer of protonated OCS in the ISM above 50\,GHz. A search for HOCS\+ towards star-forming regions at 3\,mm ($\sim$\,90\,GHz) was carried out some time ago despite the large uncertainties of the rest frequencies ($\sim$\,800 kHz, corresponding to $\sim$\,2.7\,km/s at 90\,GHz), and was ultimately unsuccessful \citep{tur90}. This illustrates the necessity for high–resolution laboratory spectroscopy, which provides frequencies to sufficient accuracy to allow identification of molecular carriers of emission features, particularly in cold, quiescent astronomical sources with very narrow line-widths. Motivated by the lack of precise data in the mm- and submm-wave range, a high--resolution laboratory study of the ground state isomer was carried out. This work has resulted in  precise measurements of HSCO\+ and DSCO\+ up to $\sim$\,400\,GHz.

\section{Experiment}

The spectroscopic study of HSCO\+ and DSCO\+ performed here has involved three different spectrometers: two at the Center for Astrochemical Studies (CAS) at the Max Planck Institute for Extraterrestrial Physics in Garching (DE), and one at the Harvard--Smithsonian Center for Astrophysics (CfA) in Cambridge (USA).\\

\subsection{The CASAC experiment}

Chronologically, the first part of the experiment was begun with the new frequency-modulated free-space absorption cell spectrometer, CASAC (the Center for Astrochemical Studies Absorption Cell), recently developed at our institute \citep{2017A&A...602A..34B}.
The main radiation source is a frequency synthesiser (Keysight E8257D), synchronised to a 10\MHz rubidium frequency standard (Stanford Research Systems) for accurate frequency and phase stabilisation. The radiation from the synthesiser is then coupled to a Virginia Diodes (VDI) solid state active multiplier chain, which allows a great frequency agility and a full coverage of the 75--1100\GHz frequency range. The radiation is fed through a Pyrex tube, 3\,m long and 5\,cm section, equipped with two hollow stainless steel electrodes, 10\,cm long, connected to a DC power supply (5\,kW). The 2\,m active region of the discharge, defined by the distance of the two electrodes, can be cooled by liquid nitrogen and magnetically confined using a 3-layer solenoidal coil wrapped around the glass tube and which can generate a magnetic field coaxial with the radiation beam up to $\sim$\,300\,G. This technique is particularly suited for producing positive molecular ions since the magnetic field increases the length of the ion rich negative glow by restricting inside a small diameter tube the ionising electrons accelerated by the large cathode drop of an anomalous glow discharge \citep{1983JChPh..78.2312D}. The frequency modulation of the radiation is obtained by encoding its signal with a sine-wave at a rate of 15\kHz; the signal, after interacting with the molecular plasma, is hence detected by a liquid-He cooled InSb hot electron Bolometer (QMC Instruments Ltd.). A lock-in amplifier (SR830, Stanford Research Systems) is used for demodulating the detector output at twice the modulation frequency ($2f$ detection) resulting in a second derivative profile of the absorption signal recorded by the computer controlled acquisition system.

The chemical conditions for the production of the protonated carbonyl sulfide were first taken from those producing the largest amount of HOCO\+, a similar and isoelectronic species recently observed in our laboratory \citep{2017A&A...602A..34B}; once the production of the latter was optimised, the CO$_2$ sample was replaced with OCS. The experimental conditions that produced the largest signal of HSCO\+ were further optimised and found to be a 1:1 mixture of OCS and H$_2$, diluted in a buffer gas of Ar, producing a total pressure, as measured at the output of the absorption cell, of 20\,mTorr; for the DSCO\+ molecule experiment, the H$_2$ was replaced by the D$_2$ sample. Further parameters that were crucial for the signal quality were a DC discharge with 5\,mA at $\sim$\,1.5kV, a magnetic field of $\sim$\,200\,G and a temperature of the glass wall of $\sim$\,130\,K. The latter has been the most critical factor: the optimal conditions were found to be varying during the experiment, more so than in other similar studies. 

\subsection{The Double-Resonance Experiment}

The CfA Fourier Transform Microwave (FTM) spectrometer was used to observe the \emph{b}-type spectrum of HSCO\+. The same instrument was used in the original detection of the protonated species \citep{mcc07} and modified accordingly to perform the double-resonance (DR) experiment; the full description of the instrument and the DR implementation can be found elsewhere \citep[e.g.][]{2000ApJS..129..611M, 2010JChPh.133s4305L}. Briefly, protonated ions are created in the throat of a small supersonic nozzle by applying a low-current dc discharge (750\,V, 0.3\,mA) to a short (300\,$\mu$s) gas pulse created by a Series 9 nozzle. The molecular beam was formed by 1.2\% of OCS in H$_2$ diluted further in a 1 to 5 ratio with pure He, reaching a total flow rate of 20\,cm$^3$\,min$^{-1}$ at standard temperature and pressure. The stagnation pressure behind the valve was 2.5\,kTorr and the valve operated with a repetition rate of 6\,Hz. The frequency coverage of the FTM spectrometer (5--43\,GHz) can be ``extended'' by means of DR measurements: the FTM cavity is kept fixed to a known resonant frequency corresponding to a specific transition (the ``probe'') of the target molecule, and, during the free induction decay of the FTM signal, a second radiation source (the ``pump'') is scanned in frequency until a resonance is reached with another transition sharing a common rotational level. Depending on the frequency range of the pump signal, microwave-microwave (MW--MW) or millimiterwave-microwave (MMW--MW) DR techniques are defined. In the search for the \emph{b-}type spectrum of HSCO\+ the MMW--MW experiment was performed coupling the output of a frequency synthesiser (Keysight N5173B-540) with a frequency tripler. The signal was then amplified (Millitech AMP-10-10040) and sent to a second frequency tripler unit (VDI Inc. WR2.8X3) to reach the frequency range of 250--375\,GHz. The nominal input power is 15\,dBm, and the final high frequency output is on the order of several dBm.

\subsection{The Free Unit Jet Experiment}

Finally, the last part of the measurements for the HSCO\+ cation was conducted with the newly developed Free Unit Jet Experiment in Garching. A deeper elucidation of this spectrometer will be the subject of a follow-up work (Lattanzi et al. in preparation), while here we will briefly report the basics of the instrumentation. A molecular beam is created by mixing the output of several mass flow controllers (MKS Instruments) that are individually connected to the sample bottles, and then injecting this mixture downstream into a vacuum chamber through the 1--mm pinhole of a pulsed valve (Series 9, Parker Hannifin). The expansion of the molecular beam into the chamber is supersonic thanks to the high pressure gradient (up to 10$^{10}$) between behind the valve (few kTorr/ few bar) and the chamber ($\sim$\,10$^{-7}$\,Torr/$\sim$\,10$^{-5}$\,bar); this supersonic expansion allows to adiabatically cool the molecular beam, reaching temperatures in the range of 7 to 20\K ca., depending on the buffer gas used. The vacuum inside the chamber is obtained by means of a large diffusion pump (DIP 8000, Oerlikon Leybold), supported by a combination of mechanical pumps (roots blower and rotary--vane pump, Oerlikon Leybold). The coupling of the molecular beam to the mm-- and submm--wave radiation is obtained through a roof--top mirror placed inside the chamber, which also contains the aperture through which the molecular sample is injected. The probing radiation enters the vacuum section through a teflon window on the opposite side of the roof--top mirror, interacts twice with the gas and finally reflects back outside the chamber. The radiation source (harmonic multiplication of the signal generated by a frequency synthesiser) and the detector (hot electron bolometer) are the same as those used for the CASAC experiment, allowing the same frequency coverage and agility. The production of unstable species is achieved by attaching a high-voltage low-current DC nozzle to the front of the valve, through which the molecules pass right after the pulsed valve and prior to free expansion, where the molecular sample is quickly stabilised in the region dubbed the \lq\lq zone of silence''. All the measurements are performed in absorption, with frequency modulation of the signal, as described above for the CASAC. \\
The experimental conditions were first optimised on the well--known protonated ion HCO\+. Finally, a highly diluted mixture of OCS in H$_2$ (0.3\%) was injected through the pulsed valve, operating at a repetition rate of 15\Hz and open for 1\,ms. The discharge nozzle was operated for 1.5\,ms at 1.5\,kV ($\lesssim$ 1\,mA current). With these conditions, the final operating pressure inside the chamber was several tens of $\mu$Torr. The continuous--wave (CW) signal of the synthesiser, synced to a 10\MHz rubidium frequency standard (Stanford Research Systems), is frequency--modulated at 30\,kHz and, after the interaction with the molecular beam, acquired by the hot electron bolometer before passing through a lock-in amplifier (MFLI, Zurich Instruments). Here the signal is demodulated at twice the modulation frequency with a 30\,$\mu$s time constant, and digitised so that integration of the molecular signal and baseline subtraction can be performed to yield the final spectrum. All the timings (discharge, lock-in demodulation, and frequency stepping) were triggered to the repetition rate of the valve.

\section{Analysis}

\begin{figure}[htbp]
\includegraphics[width=\columnwidth]{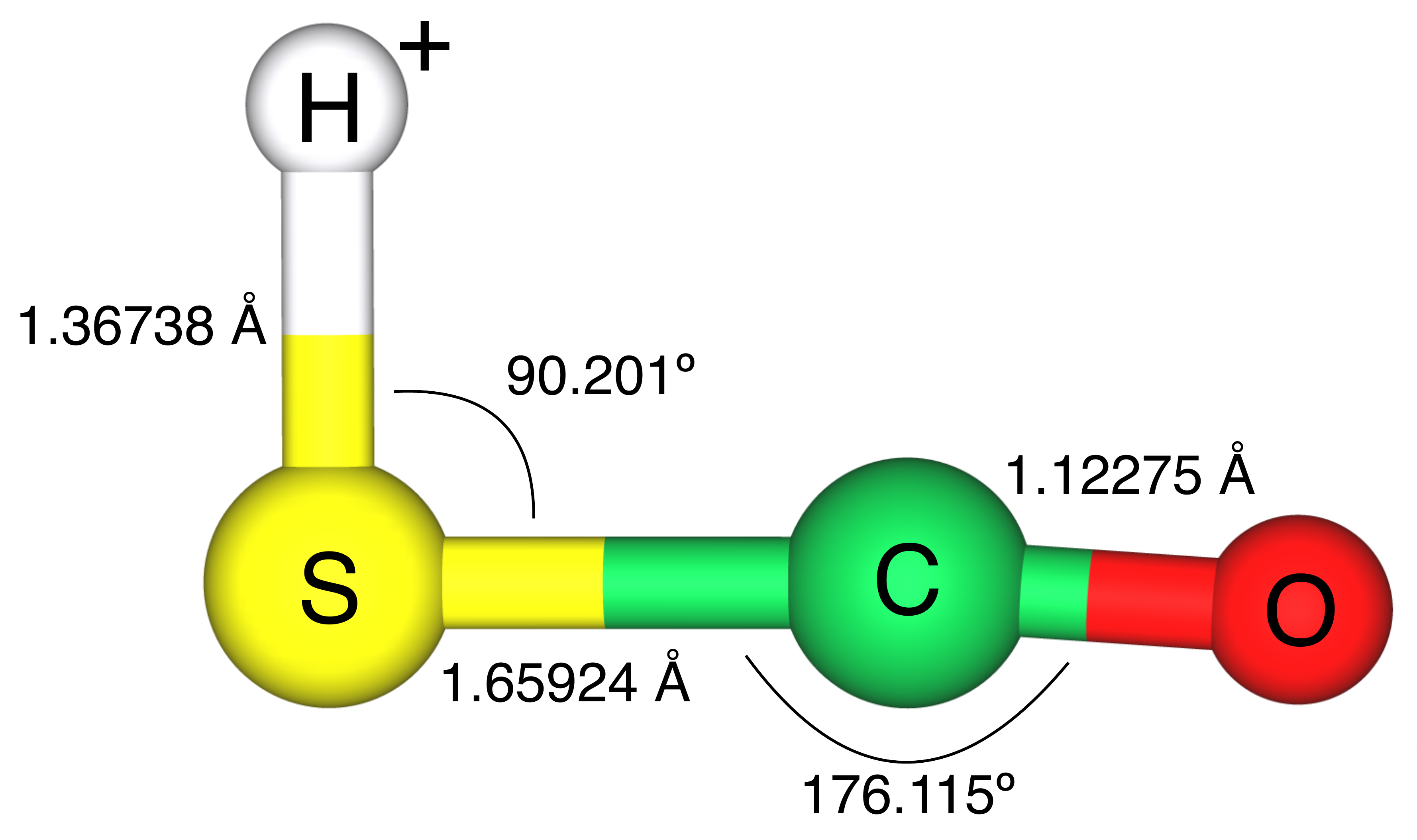}
\caption{Theoretical geometry of protonated carbonyl sulfide determined by \citet{2012JPCA..116.9582F}.}
\label{fig:1}
\end{figure}

The protonated carbonyl sulfide system includes two stable isomeric forms, HSCO\+ and HOCS\+, separated by 4.9\,kcal/mol. The two isomers are connected by a transition state with the proton localised above the central carbon atom, lying 68.9\,kcal/mol above the ground state HSCO\+ \citep{whe06}. The latter is a closed-shell near-prolate ($\kappa=-0.992$) asymmetric top and exhibits both \emph{a-} and \emph{b-}type rotational spectra ($\mu_a = 1.57$\,D and $\mu_b = 1.18$\,D). The OCS angle is nearly linear at approximately 175$^{\circ}$ (Fig. \ref{fig:1}).\\

The search for HSCO\+ in the millimeter band was guided by the rotational constants previously reported by \citet{mcc07} in the cm-wave band along with the centrifugal distortion constants derived by \citet{2012JPCA..116.9582F}. The low--frequency measurements detected the three lowest $K_a$=0 rotational transitions which provided a constrained $B_{eff}$ (= $[B+C]/2$) constant. $B_{eff}$ and the predicted quartic distortion constant $D_J$ were subsequently used in the initial search to observe the $a-$type rotational transition $27_{0,27}-26_{0,26}$ near 304\GHz. In the first 50\MHz--wide search we found just one candidate feature, deviating only by $\sim$\,2\MHz from the prediction, that fulfilled all the desirable properties: it (1) was produced only in presence of a DC discharge, (2) was visible only when a magnetic field was applied, and (3) disappeared when either H$_2$ or OCS was removed from the sample mixture. The decisive test for confirming the assignment was then to search the harmonic progression of the rotational transitions, and finally we were able to detect a total of 11 $K_a=0$ lines, all exhibiting the same experimental behaviours. The identification of lines of HSCO\+ in the millimeter band was then supported by the detection of additional 33 rotational transitions in the $K_a$ ladders, up to $K_a=6$, that were close to the prediction. \\
Despite the addition of these newly detected rotational transitions, the uncertainty on the $A$ rotational constant remained large, as expected considering that the new set of millimeter/submillimeter lines all belonged to the \emph{a}-type spectrum of the ion, and were hence less sensitive to the contribution of the \emph{A} constant. The uncertainty on the aforementioned parameter was around 20\,MHz, compared to those for the \emph{B} and \emph{C} rotational constants, which were on the order of a few kHz. With the parameters derived from the \emph{a}-type spectrum analysis, a search with the CASAC experiment for the \emph{b}-type spectrum of HSCO\+ was then carried out. This search, however, led to a non detection, owing to both a large uncertainty on the predicted frequencies and a weaker dipole moment, which results roughly in lines expected with half intensities compared to the \emph{a}-type spectrum.

\begin{figure}[htbp]
\includegraphics[width=\columnwidth]{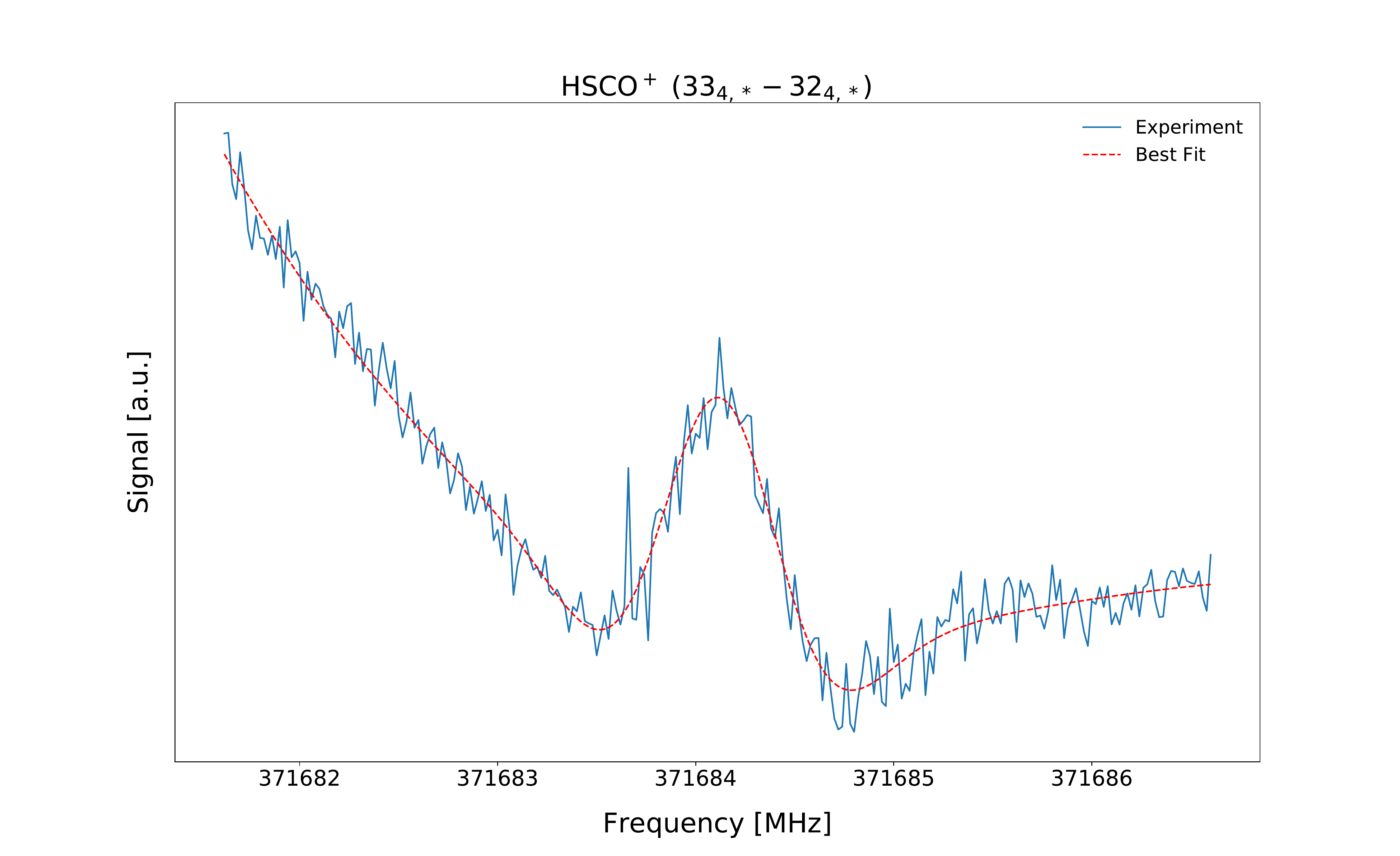}
\includegraphics[width=\columnwidth]{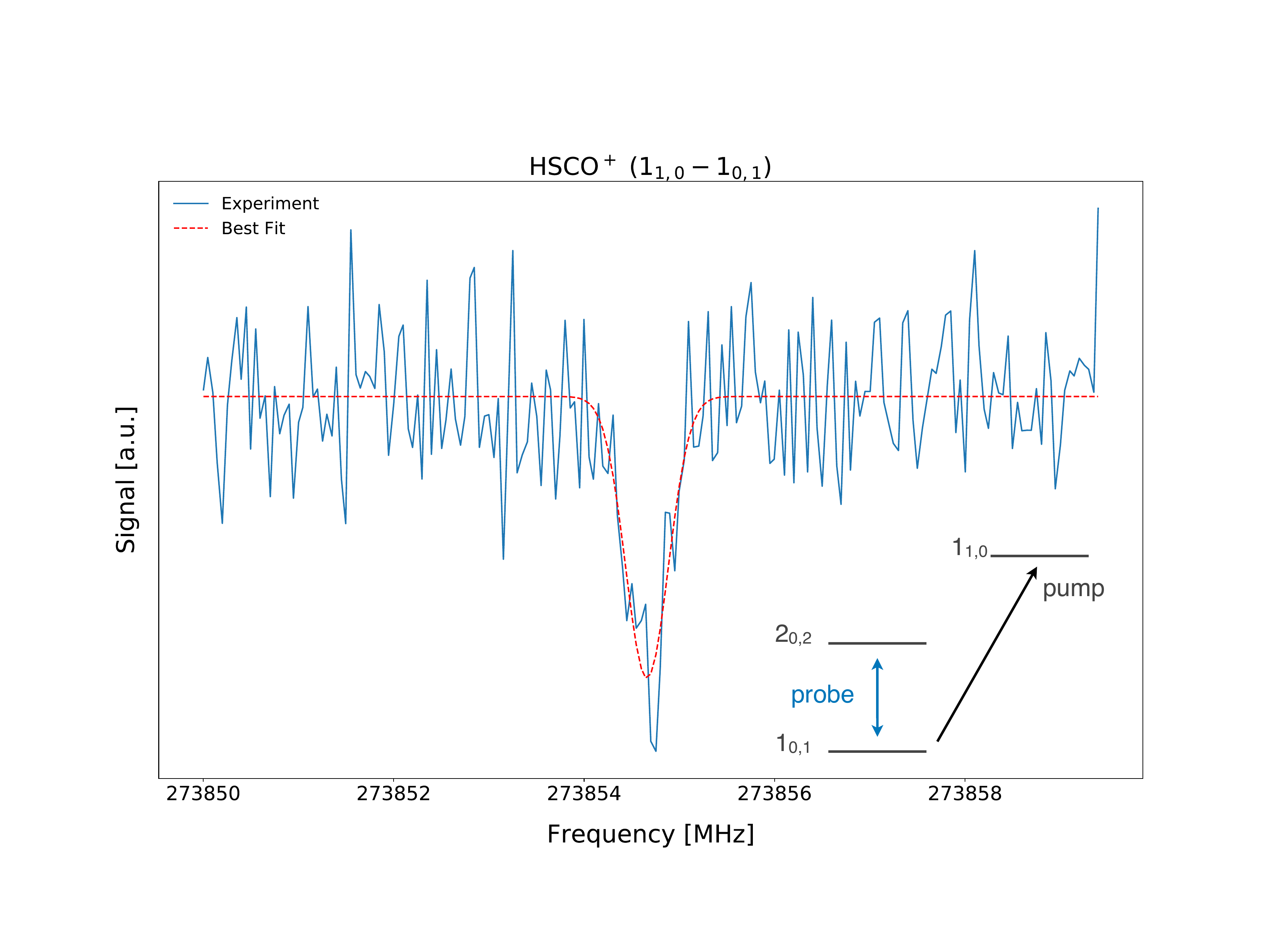}
\caption{Laboratory spectra of HSCO\+: (\emph{top}) the $J_{K_a,K_c}=33_{4,*}-30_{4,*}$ (overlapping) doublet around 371\,GHz, acquired in 725\,s integration time with 3\,ms time constant; the strong spikes visible in plot are due to instabilities of the dc discharge. (\emph{bottom}) The \emph{b}-type $J_{K_a,K_c}=1_{1,0}-1_{0,1}$ rotational transition around 274\,GHz, acquired in double resonance with 45\,kHz step and in a total integration time of 20 minutes (30 shots per integration step at 5\,Hz). The red dashed lines represent the best fit to a speed-dependent Voigt and Gaussian profile (respectively). A schematic diagram indicating the two transitions that share a common energy level is shown in the inset.}
\label{fig:2}
\end{figure}

The \emph{b}-type spectrum was then searched for using the MMW--MW double resonance techniques. The strongest depletion of the probe transition, and hence the largest DR signal, is achieved when the pump rotational transition is connected to the lowest energy state of the probe rotational transition; also for this reason the best option was to look for the fundamental $b-$type transition 1$_{1,1}$--0$_{0,0}$, which was predicted around 285\GHz and whose corresponding $a-$type transition 1$_{0,1}$--0$_{0,0}$ were previously detected in the same FTMW spectrometer. After a few searches around the predicted value, a clear DR signal was finally detected at a frequency $\sim$\,140\,\MHz away from the prediction (0.05\%). Once the fundamental rotational $b-$type was detected, and hence the $A$ rotational constant was locked to its value with a reasonable uncertainty, the search for the other transitions of the same spectrum, detectable within our spectral coverage, was straightforward and we were able to detect a total of 5 $b-$type transitions (e.g. Fig.\ref{fig:2}). 

\begin{figure}[htbp]
\includegraphics[width=\columnwidth]{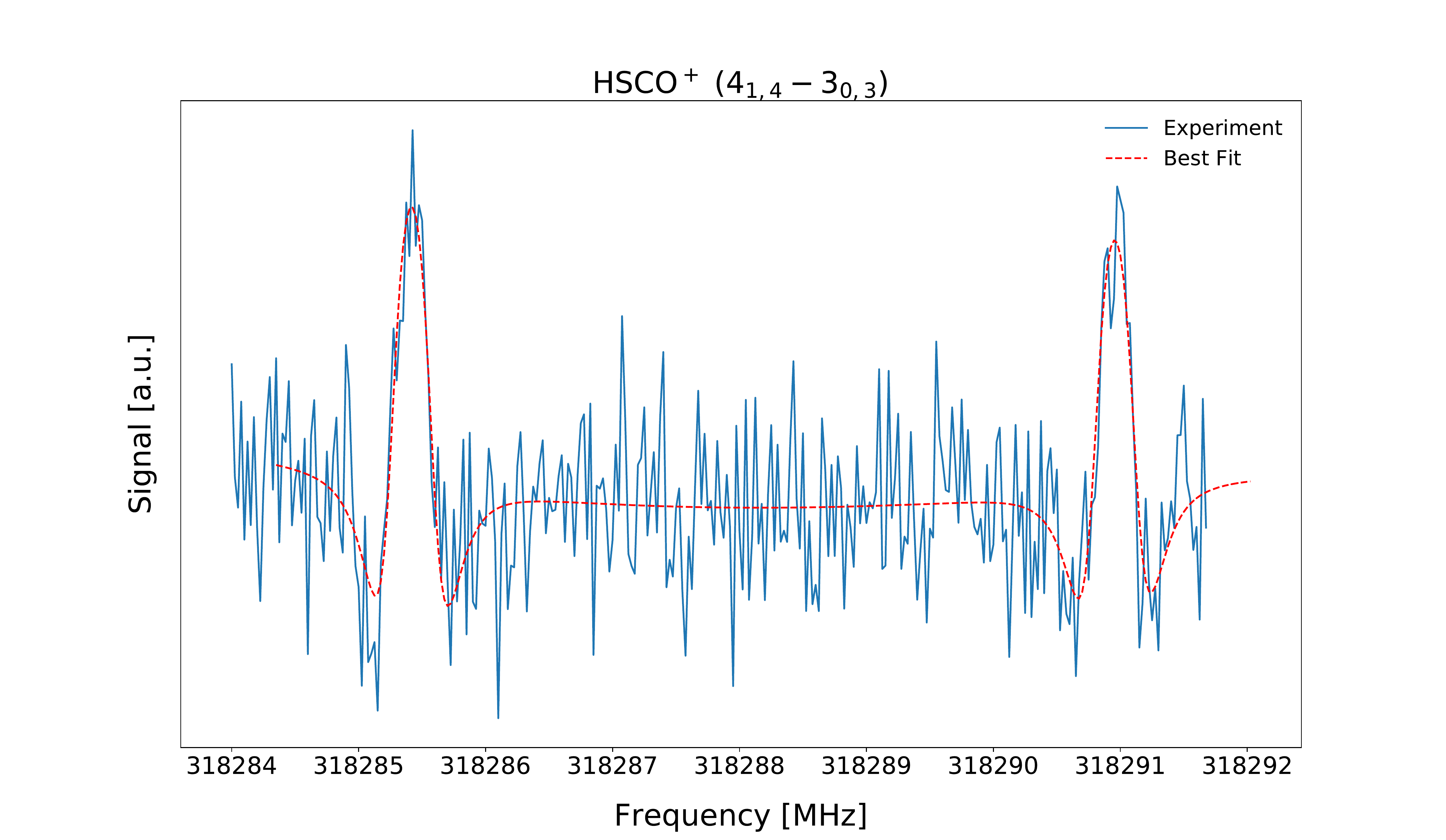}
\caption{Laboratory spectrum of the HSCO\+ $J_{K_a,K_c}=4_{1,4}-3_{0,3}$ rotational transition around 318\,GHz, acquired with the free--unit jet experiment. The integration time is $\sim$\,15 minutes with 30\,$\mu$s time constant. The red dashed line represents the best fit to a speed-dependent Voigt profile, as discussed in the text. Each rotational transition has a double--peaked line shape, the result of the Doppler shift of the supersonic molecular beam relative to the two traveling waves that compose the radiation beam.}
\label{fig:3}
\end{figure}

With the DR measurements as a guide, we decided to test our newly developed free--unit jet experiment with the search of other low--energy $b-$type HSCO\+ rotational transitions. The experimental conditions described above allowed the detection of a total of 16 $b-$type features, including the line previously observed with the DR experiment, belonging to $Q-$ and $P-$branches between 274 and 373\,\GHz (e.g. Fig.\ref{fig:3}). 

The rotational spectrum of HSCO\+ was analysed with the Watson S-reduced Hamiltonian, including all the quartic centrifugal distortion terms and two sextic terms, $H_{JK}$ and $H_{KJ}$. The quartic parameter $D_K$ was not constrained in our dataset due to a strong correlation with the $A$ rotational constant; the overall rms improved slightly ($\lesssim$\,10\%) by keeping this parameter fixed to the \emph{ab initio} value. All 63 rotational transitions, including the three observed in the previous microwave experiment, were reproduced by our model with a final rms uncertainty of 39\,kHz, which corresponds approximately to the average experimental uncertainty (Table \ref{table:1}). 

Using the absorption cell, we employed a similar experimental approach for studies of DSCO\+, replacing hydrogen with deuterium. The experimental search started with the hunt for the $29_{0,29}-28_{0,28}$ line, as the prediction from the microwave measurements estimated this line around 318\,GHz. In a matter of a few days we collected 50 $a-$type rotational transitions, up to the $K_a=7$ ladders. With the same Hamiltonian adopted for the hydrogen protonated ion and using the same set of molecular parameters, the measured DSCO\+ rotational spectrum was fitted to an rms uncertainty of 23\,kHz  (Table \ref{table:2}). \\ 
All the experimental lines, except for the DR spectra, were fitted using the line profile model summarised in \citet{2003JMoSp.221...93D} and implemented in our in-house analysis software. More precisely, a speed-dependent Voigt profile was applied to retrieve the central frequency of the 2$f$ absorption line; both the complex component of the Fourier-transform of the dipole correlation function (i.e. the dispersion term) and a third--order polynomial were also taken into account to model the line asymmetry and baseline produced by the background standing-waves between non-perfectly transmitting windows of the absorption cell in the CASAC measurements. The experimental uncertainty is estimated to be in the range of 30-50\,kHz, depending on the line width, the achieved signal-to-noise ratio (S/N), and the baseline.

In the supplementary material, available at the CDS, the full list of assigned transitions is available for both HSCO\+ and DSCO\+, along with the respective frequencies and estimated uncertainties.

\section{Discussion and Conclusions}

A combination of spectroscopic techniques has been used to extend the frequency coverage of the protonated OCS system, a cation of potential radioastronomical interest. This new set of experiments allow the detection of 60 and 50 new rotational transitions for HSCO\+ and DSCO\+, respectively. First detection of $b-$type transitions for HSCO\+ permitted the \emph{A} rotational constant to be precisely determined, free from correlations with other spectroscopic parameters. The experimental rotational constants to {ab initio} values are in striking agreement: both \emph{B} and \emph{C} agree to 0.01\%, while \emph{A} only differs by 0.07\% (Table\,\ref{table:1}). The agreement of the experimental and theoretical quartic centrifugal distortion terms is in the range of a few percent, while the agreement is better than 25\% for the capital $D'$s and lower $d$'s, respectively. This level of agreement is similar to that obtained with the same level of theory (CcCR QFF) recently reported for HOCO\+ \citep{2017A&A...602A..34B}. For example, the experimental $D_J$ fourth-order distortion constant differs by 2.5\% from the \emph{ab-initio} values; recent studies on rigid molecules with two to three heavy atoms showed that this parameter, when calculated at a high level of theory, is usually accurate to about 3\% or better \citep{2011A&A...533L..11L, 2012ApJS..200....1S}. In Table\,\ref{table:1} the HSCO\+ spectroscopic parameters are also compared with those derived by \citet{0004-637X-706-2-1588} for HSCN. This parallel confirms, once again, (1) how close are the rotational constant values and their centrifugal distortion corrections for two isoelectronic species; (2) how the non-detection of the $b-$type transitions \citep[as in the work by][]{0004-637X-706-2-1588} affects the accuracy of the \emph{A} rotational constant. A similar level of agreement between theory and experiment is found for the spectroscopic parameters of the deuterated species. However the \emph{A} rotational constant needs to be treated more carefully, since the $b-$type spectrum is presently lacking for this species (Table\,\ref{table:2}).\\

\begin{figure}[htbp]
\includegraphics[width=\columnwidth]{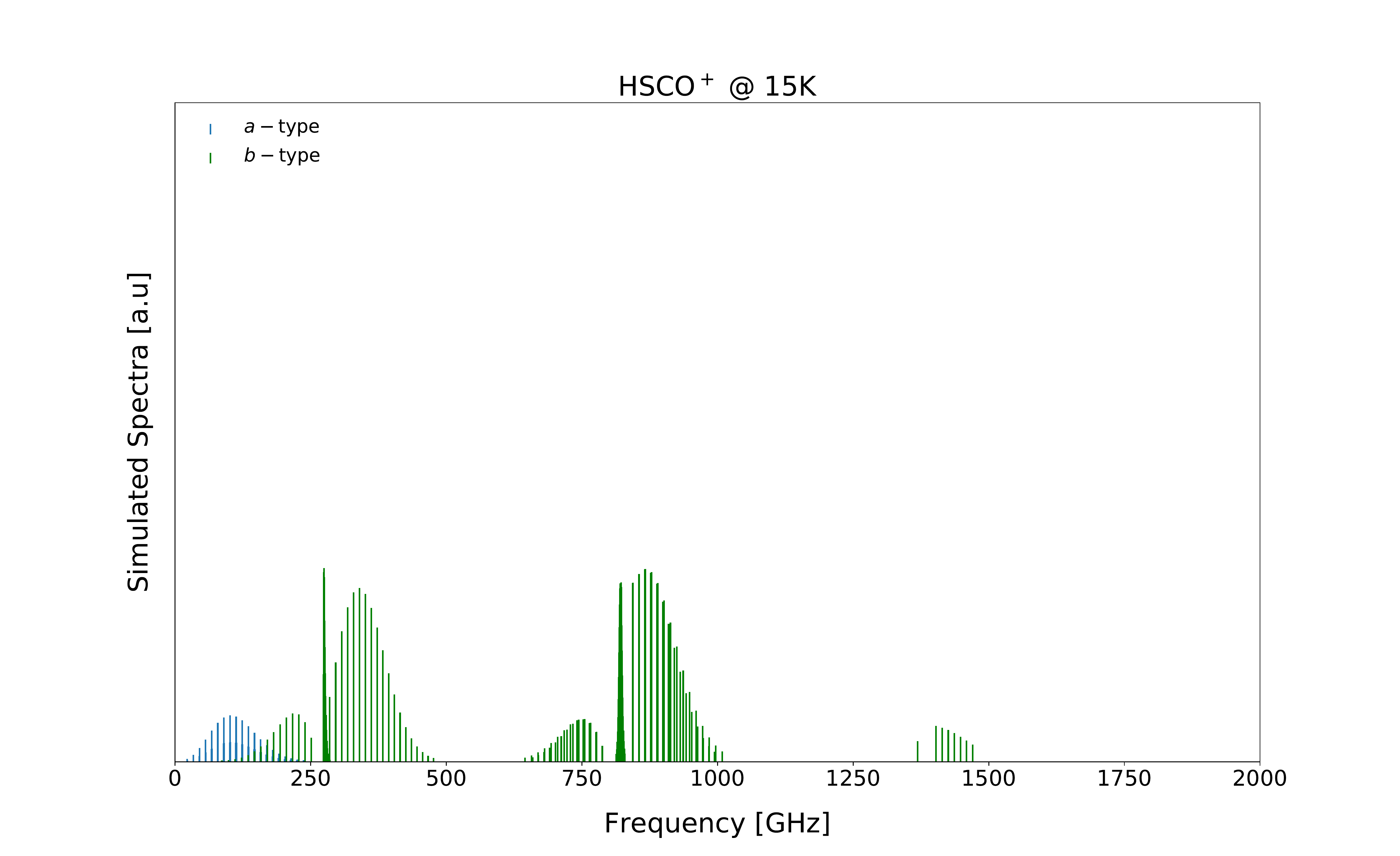}
\includegraphics[width=\columnwidth]{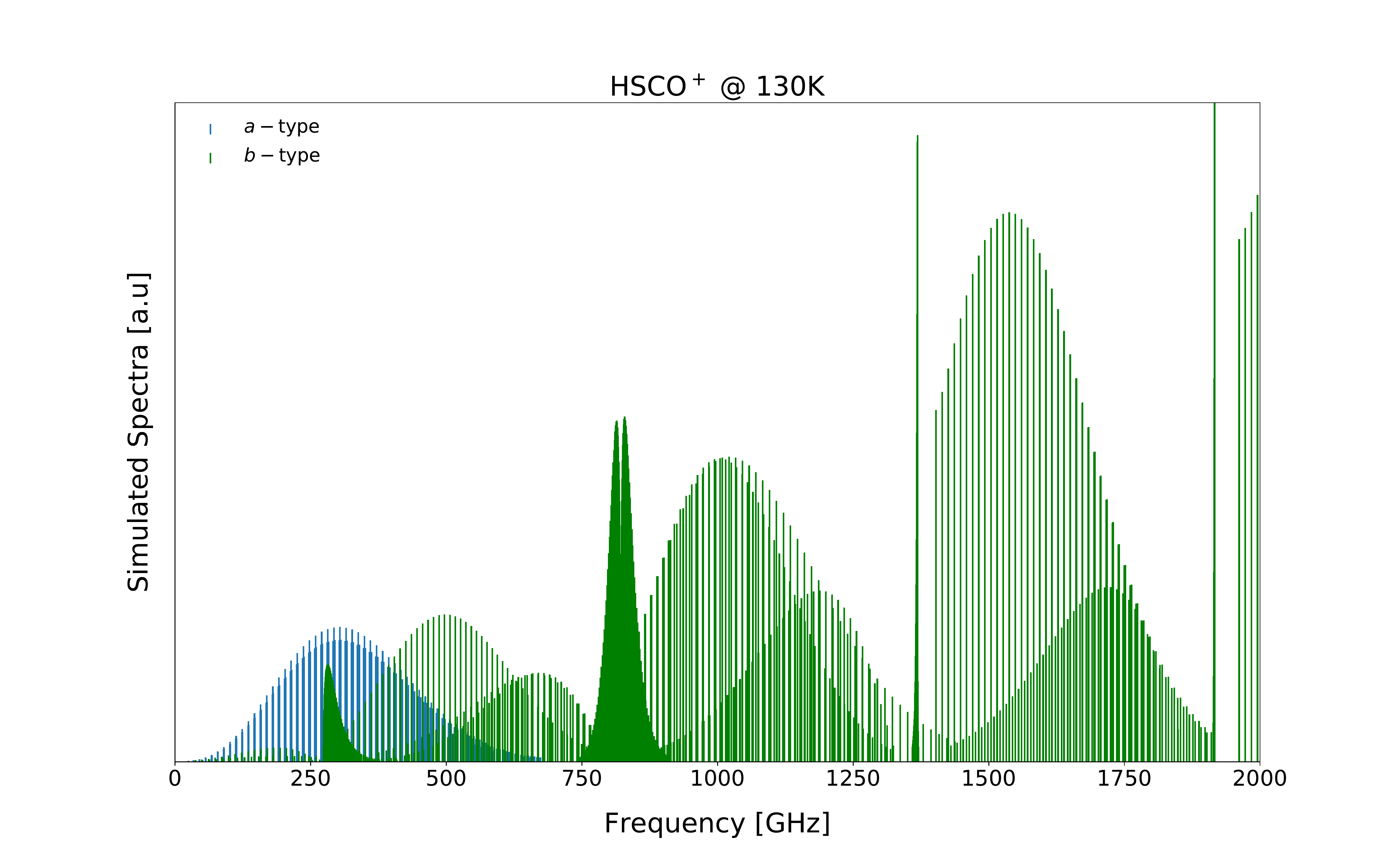}
\caption{Simulated spectra of HSCO\+. Thermal conditions were chosen representing the average temperatures of the Free--Unit Jet (\emph{top}) and CASAC (\emph{bottom}) experiment, respectively. The $y$ axis scale is the same for both plots.}
\label{fig:4}
\end{figure}

The present work illustrates the importance of using different but complementary techniques to spectroscopically characterise a protonated species at high frequency. The CASAC experiment addressed the \lq\lq warmer'' part of the spectrum, extending the microwave measurements to considerably higher $J$ (up to $J = 36$) and $K_a$ (up to $K_a = 6$) levels. However, attempts to address other aspects of the rotational spectrum with this experiment were unsuccessful, for several reasons. Although moving to a lower frequency range provided more power from the radiation source, the targeted transitions sought are poorly populated in the relatively warm molecular plasma of the CASAC ($\sim$\,130\,K). Attempts to enhance the population of these lower states by decreasing the temperature of the absorption cell resulted in the freeze--out of the OCS sample on the glass wall and an overall decrease of the absorption signal.
Extension to higher frequency also proved impractical because the power of the millimeter--wave source slowly decreases with increasing frequency, and the temperature of the plasma is no longer optimal to probe this higher--lying transitions. These considerations only apply to the $a-$type spectrum of the ion, which has a linear-like progression with increasing frequency, similar to that of strictly linear molecules. The $b-$type spectrum has a more complex progression but provides access to a wide energetic regimes (Figure\,\ref{fig:4}). Owing to a combination of the large uncertainty of the predicted $b-$type transitions and the technical challenge of stabilising the discharge system for a longer integration time required for this search, no $b-$type absorption features were observed in the CASAC experiment.  Despite the intrinsic limitations of finding connected transitions sharing the same lower energy rotational level, DR experiments are generally unambiguous, since perturbations to the probe transition can only arise when the pump frequency is coincident with a connected transition. In addition, this experiment greatly benefits from the use of the same spectrometer where HSCO\+ was already observed, meaning that the experimental conditions were already optimised.\\

With the rotational constants reasonably well constrained from a combination of CASAC and DR measurements, the spectroscopic analysis was extended even further by detection of 16 additional $b-$type rotational transitions using a new jet experiment in Garching. This new jet experiment uses a similar nozzle source to that routinely employed in the FTM experiments (i.e. a free expansion of a pulsed supersonic molecular jet) and, although less sensitive than the DR implemented in the FTM spectrometer, the former has notable advantages in terms of frequency agility, and flexibility with respect to rotational temperature of the expansion. Its wide frequency coverage permits the detection of rotational lines through the millimeter and sub-millimeter molecular spectra, with the only limitation being the choice of the buffer gas used for the jet expansion, and hence the resulting effective rotational temperature. For this experiment, the choice of buffer gas is fixed, namely H$_2$, for protonated species. Owing to the large $A$ rotational constant, these conditions limit the measured lines to $K_a = 0\rightarrow 1$ transitions; attempts to detect transitions with higher combination of $J/K_a$ failed presumably because of the low population of higher energy rotational levels in the jet expansion. Nevertheless, the spectroscopic catalogues for HSCO\+ and DSCO\+ are largely improved, now allowing radioastronomical search to be performed with much greater confidence.  With the new spectroscopic data, the rest frequencies of the astronomically most interesting lines have either been measured or can be predicted to better than 100\,kHz for both HSCO\+ and DSCO\+ up to 500\,\GHz, equivalent to better than 60 m/s in terms of equivalent radial velocity.

\begin{acknowledgements}

The authors wish to thank Mr. Christian Deysenroth and Mr. Martin Gillhuber for the thorough assistance in the engineering of the molecular spectroscopy laboratories at the MPE/Garching. We would also like to thank Edward Tong of the Submillimeter Array Receiver laboratories at the Harvard-Smithsonian Center for Astrophysics for loaning the Millitech amplifier used for the 300 GHz DR measurements. M.C.\,McCarthy and K.L.K.\,Lee thank NSF grant AST-1615847 for financial support.

\end{acknowledgements}

\bibliographystyle{aa}
\bibliography{references.bib}


\clearpage

\begin{table}
\caption{Spectroscopic parameters derived for HSCO\+}
\label{table:1} \centering
\begin{tabular}{llr@{.}lr@{.}lr@{.}lr@{.}l}

\hline\hline   
\noalign{\smallskip}
Parameter & unit & \multicolumn{2}{c}{This work} & \multicolumn{2}{c}{Previous\tablefootmark{a}}  & \multicolumn{2}{c}{\emph{ab initio}\tablefootmark{b}} & \multicolumn{2}{c}{HSCN\tablefootmark{c}} \\
\hline
\noalign{\smallskip}
   $A$      & \MHz  &   279431&995(10)                &   \multicolumn{2}{c}{--}           &   279236&6                       &   \multicolumn{2}{l}{289737(64)} \\
   $B$      & \MHz  &     5696&74645(86)              &     5636&8660(20)\tablefootmark{d} &     5697&6                       &   5794&71368(20) \\
   $C$      & \MHz  &     5576&97793(66)              &     5636&8660(20)\tablefootmark{d} &     5577&6                       &   5674&93940(20) \\
   $D_J$    & \kHz  &        1&51961(25)              &        3&10(10)\tablefootmark{d}   &        1&481                     &   1&66557(21)  \\
   $D_{JK}$ & \kHz  &      136&393(93)                &   \multicolumn{2}{c}{--}           &        \multicolumn{2}{c}{137}   &   151&53(11) \\ 
   $D_K$    & \MHz  &       15&654\tablefootmark{e}   &   \multicolumn{2}{c}{--}           &       15&654                     &   \multicolumn{2}{c}{--}  \\
   $d_1$    & \Hz   &      -34&03(17)                 &   \multicolumn{2}{c}{--}           &       -30&14                     &   -35&91(21)\\
   $d_2$    & \Hz   &       -4&720(85)                &   \multicolumn{2}{c}{--}           &       -3&829                     &   -5&17(31)\\
   $H_{JK}$ & \Hz   &        0&320(49)                &   \multicolumn{2}{c}{--}           &        0&361                     &   0&518(26)\\
   $H_{KJ}$ & \Hz   &     -176&9(10)                  &   \multicolumn{2}{c}{--}           &       -180&366                   &   -170&3(86)\\
   $H_K$    & \kHz  &        0&64769\tablefootmark{e} &   \multicolumn{2}{c}{--}           &       0&64769                    &   \multicolumn{2}{c}{--}   \\

\hline
\noalign{\smallskip}
\# lines       &      & \multicolumn{2}{c}{63} \\
$\sigma_{rms}$ & \kHz & \multicolumn{2}{c}{41} \\  
$\sigma_{w}  $\tablefootmark{f} &      &      1&19              \\  
\hline
\end{tabular}
\tablefoot{Values in parentheses represent 1$\sigma$ uncertainties, expressed in units of the last quoted digit.\\
\tablefoottext{a}{\citet{mcc07}.}\\
\tablefoottext{b}{\citet{2012JPCA..116.9582F}.}\\
\tablefoottext{c}{\citet{0004-637X-706-2-1588}.}\\
\tablefoottext{d}{The actual parameters fitted in \citet{mcc07} are $B_{eff}=(B+C)/2$ and $D_{eff}=D_J+(B-C)^2/\{32[A-(B+C)/2]\}$, assuming $D_J=1$\kHz.}\\
\tablefoottext{e}{Fixed to the \emph{ab initio} value.}\\
\tablefoottext{f}{Dimensionless \emph{rms}, defined as $\sigma_{w} = \sqrt{\frac{\sum_i\left(\frac{\delta_i}{err_i}\right)^2}{N}}$, where the $\delta$'s are the residuals weighted by the experimental uncertainty (\emph{err}) and \emph{N} the total number of transitions analysed.}}
\end{table}


\begin{table}
\caption{Spectroscopic parameters derived for DSCO\+}
\label{table:2} \centering
\begin{tabular}{llr@{.}lr@{.}lr@{.}l}

\hline\hline   
\noalign{\smallskip}
Parameter & unit & \multicolumn{2}{c}{This work} & \multicolumn{2}{c}{Previous\tablefootmark{a}}  & \multicolumn{2}{c}{\emph{ab initio}\tablefootmark{b}} \\
\hline
\noalign{\smallskip}
   $A$      & \MHz  &   145158&6(22)                  &   \multicolumn{2}{c}{--}      &   145002&7                          \\
   $B$      & \MHz  &     5618&1838(19)               &     5509&6970(20)\tablefootmark{c} &     5619&3                          \\
   $C$      & \MHz  &     5401&1678(19)               &     5509&6970(20)\tablefootmark{c} &     5402&3                          \\
   $D_J$    & \kHz  &        1&42281(27)              &       12&00(10)\tablefootmark{c}   &        1&387                        \\
   $D_{JK}$ & \kHz  &        132&643(86)             &   \multicolumn{2}{c}{--}      &        \multicolumn{2}{c}{132}                        \\ 
   $D_K$    & \MHz  &       3&951\tablefootmark{d}   &   \multicolumn{2}{c}{--}      &       3&951\\
   $d_1$    & \kHz  &       -0&05913(48)              &   \multicolumn{2}{c}{--}      &       -0&05318                      \\
   $d_2$    & \kHz   &      -0&01525(11)              &   \multicolumn{2}{c}{--}      &       -0&01302                      \\
   $H_{JK}$ & \Hz   &        0&439(52)                &   \multicolumn{2}{c}{--}      &       \multicolumn{2}{c}{--}   \\
   $H_{KJ}$ & \Hz  &       -42&68(81)              &   \multicolumn{2}{c}{--}      &        \multicolumn{2}{c}{--}   \\
\hline
\noalign{\smallskip}
\# lines       &      & \multicolumn{2}{c}{53} \\
$\sigma_{rms}$ & \kHz & \multicolumn{2}{c}{23} \\  
$\sigma_{w}  $\tablefootmark{e} &      &      0&74              \\  
\hline
\end{tabular}
\tablefoot{Values in parentheses represent 1$\sigma$ uncertainties, expressed in units of the last quoted digit.\\
\tablefoottext{a}{\citet{mcc07}.}\\
\tablefoottext{b}{\citet{2012JPCA..116.9582F}.}\\
\tablefoottext{c}{The actual parameters fitted in \citet{mcc07} are $B_{eff}=(B+C)/2$ and $D_{eff}=D_J+(B-C)^2/\{32[A-(B+C)/2]\}$, assuming $D_J=1$\kHz.}\\
\tablefoottext{d}{Fixed to the \emph{ab initio} value.}\\
\tablefoottext{e}{Dimensionless \emph{rms}, defined as $\sigma_{w} = \sqrt{\frac{\sum_i\left(\frac{\delta_i}{err_i}\right)^2}{N}}$, where the $\delta$'s are the residuals weighted by the experimental uncertainty (\emph{err}) and \emph{N} the total number of transitions analysed.}
}
\end{table}

\clearpage


\end{document}